\pdfoutput=1

\documentclass[11pt]{article}

\usepackage{ACL2023}

\usepackage{times}
\usepackage{latexsym}

\usepackage[T1]{fontenc}

\usepackage{amssymb}
\usepackage{pifont}
\newcommand{\cmark}{\text{\ding{51}}}
\newcommand{\xmark}{\text{\ding{55}}}

\usepackage[utf8]{inputenc}

\usepackage{microtype}

\usepackage{inconsolata}
\usepackage{booktabs}
\usepackage{pgfplots}
\pgfplotsset{compat=1.8}
\usepgfplotslibrary{statistics,groupplots,dateplot}
\DeclareUnicodeCharacter{2212}{−}
\usetikzlibrary{patterns,shapes.arrows}
\pgfplotsset{compat=newest}

\usepackage{amsmath}

\definecolor{midnightgreen}{rgb}{0.0, 0.29, 0.33}

\definecolor{beatpurple}{HTML}{C69EC5}

\title{Dwell in the Beginning: How Language Models Embed \\ Long Documents for Dense Retrieval}

\author{ \bf
João Coelho$^{a,b}$, 
Bruno Martins$^b$,  
João Magalhães$^c$,   
Jamie Callan$^a$, 
Chenyan Xiong$^a$\\
$^a$ Language Technologies Institute, Carnegie Mellon University, United States \\
$^b$ Instituto Superior Técnico and INESC-ID, University of Lisbon, Portugal\\
$^c$ NOVA LINCS, NOVA School of Science and Technology, Portugal \\
\normalsize{jmcoelho@andrew.cmu.edu}
}
\begin{document}
\maketitle

\begin{abstract}

This study investigates the existence of positional biases in Transformer-based language models for text representation learning, particularly in the context of web document retrieval. We build on previous research that demonstrated loss of information in the middle of input sequences for causal language models, extending it to the domain of embedding learning. We examine positional biases at multiple stages of the training pipeline for an encoder-decoder neural retrieval model, namely language model pre-training, contrastive pre-training, and contrastive fine-tuning. Experiments with the MS-MARCO document collection reveal that after contrastive pre-training the model already generates embeddings that better capture the beginning of the input content, with fine-tuning further aggravating this effect.

\end{abstract}



\section{Introduction}

Recent advancements have allowed Transformer-based models to handle increasingly larger context lengths, resulting in the availability of Language Models (LMs) that can accommodate input lengths reaching tens of thousands of tokens~\citep{DBLP:journals/corr/abs-2309-16039}. However, studies assessing how well this context is captured by causal LMs~\citep{DBLP:journals/corr/abs-2307-03172} have shown that models are biased to  information contained at the beginning or end of the input, losing information in the middle. 

Instead of further analysing text generation, we extend this type of study to text representation learning, which has been a fundamental task for dense retrieval~\citep{DBLP:conf/iclr/XiongXLTLBAO21,DBLP:conf/emnlp/KarpukhinOMLWEC20}, and is also gaining attention in the context of retrieval-augmented generation~\citep{DBLP:conf/emnlp/ChevalierWAC23, DBLP:journals/corr/abs-2304-08467} and recommendation systems~\citep{DBLP:journals/corr/abs-2401-04858}. Specifically, we focus on web document retrieval, examining how well a single embedding represents a complete web document, while assessing the emergence of eventual position biases.


    

We start by continuously pre-training and fine-tuning an encoder-decoder model similar to T5-base~\citep{JMLR:v21:20-074} but with a context length of 2048 tokens, following standard techniques to achieve a model that is representative of the state-of-the-art among the low-parameter scale. We leverage the MS-MARCO (v1) document collection~\citep{DBLP:conf/nips/NguyenRSGTMD16}, as this dataset is commonly used in retrieval evaluation benchmarks~\citep{thakur2021beir, DBLP:conf/eacl/MuennighoffTMR23}, and it is one of the major sources of training data for the fine-tuning of neural retrieval models~\citep{DBLP:journals/corr/abs-2310-08232, DBLP:journals/corr/abs-2212-03533}. 


We found the existence of a \textit{dwell in the beginning} effect, i.e. a positional bias displayed by the model where earlier parts of the input are dominant in the embedding. We track this behavior by evaluating the model on position-aware tasks during multiple stages of its training. From our experiments, we conclude that these positional biases start emerging during unsupervised contrastive pre-training, and that the heavy reliance on MS-MARCO data for fine-tuning will exacerbate this behavior. Our models and code are available in a public GitHub repository\footnote{\url{https://github.com/cxcscmu/LongEmbeddingAnalysis}}.


\section{Related Work}



Bi-encoders are now the state of the art approach to dense retrieval~\citep{DBLP:conf/iclr/XiongXLTLBAO21,DBLP:conf/emnlp/KarpukhinOMLWEC20}.
Current standard training setups leverage the usage of contrastive loss functions and methods such as ANCE~\citep{DBLP:conf/iclr/XiongXLTLBAO21} to sample hard negative examples. Other techniques that are often employed include in-domain pre-training~\citep{DBLP:conf/acl/GaoC22} and retrieval-aligned pre-training~\citep{lu-etal-2021-less, DBLP:conf/emnlp/XiaoLSC22, DBLP:conf/acl/LeeCT19, DBLP:conf/sigir/MaGZFC22, DBLP:journals/corr/abs-2401-11248}, which allow for a better fine-tuning starting point, consequently achieving stronger retrieval results.


For long document retrieval, early methods dealt with the increased input length through heuristic aggregation strategies, which rely on segmenting the document into passages that are scored independently, with max-pooling being particularly effective~\citep{DBLP:conf/sigir/DaiC19}. Instead of aggregating scores, studies like PARADE~\citep{DBLP:journals/corr/abs-2008-09093} considered the aggregation of passage-level representations. Other authors~\citep{DBLP:journals/corr/abs-2207-01262} used Transformer architectures with sparse attention patterns~\citep{DBLP:journals/corr/abs-2004-05150, DBLP:conf/nips/ZaheerGDAAOPRWY20} to model the long inputs more efficiently, showing that, on MS-MARCO, the gains that arise from using such models are limited when compared to simple aggregation strategies.

Currently, LLaRA~\citep{DBLP:journals/corr/abs-2312-15503} achieves state-of-the-art performance in the MS-MARCO document retrieval task, by continually pre-training LLaMA-7B~\citep{DBLP:journals/corr/abs-2302-13971} with a retrieval-aligned task. Models like LLaRA leverage context windows of up to 4096 tokens, relying on FlashAttention~\citep{DBLP:conf/nips/DaoFERR22, DBLP:journals/corr/abs-2307-08691} for fast and exact full attention computation, together with some variation of Rotary Position Embeddings (RoPE)~\citep{DBLP:journals/ijon/SuALPBL24} or Attention with Linear Biases (ALiBi)~\citep{DBLP:conf/iclr/PressSL22}. This enables stronger modeling of longer sequences, without the need of additional training, while resorting to full-attention computations.

\section{Methodology}


This section details the training of a T5-base retriever with 2048 input length (T5-2K), adapting the T5 architecture to follow recent advancements in long-context language modeling, and following a state-of-the-art dense retrieval training pipeline. 

\subsection{Model Architecture}

We use the T5-base architecture as a backbone, replacing the positional embeddings by RoPE~\citep{DBLP:journals/ijon/SuALPBL24}. This change was motivated by RoPE's ability to extrapolate to larger contexts, and its compatibility with FlashAttention. Specifically, we use Dynamic NTK-RoPE~\citep{DBLP:journals/corr/abs-2309-00071}, which in theory allows for extrapolation to longer input sequences without further training. The retriever follows a tied bi-encoder architecture, i.e., the same model encodes both queries and documents. The T5 decoder is used as a pooler~\citep{DBLP:conf/acl/NiACMHCY22}, generating a single token and considering its representation as the document embedding.



\subsection{Dense Retriever Training Pipeline}

\noindent \textbf{Language Modelling Pre-training:} Starting from T5-base available at HuggingFace\footnote{\url{https://huggingface.co/t5-base}}, we continuously pre-train the model on 8 billion tokens from the MS-MARCO document collection, for the model to adapt to the new maximum sequence length, new positional embeddings, and MS-MARCO's document distribution. We follow the original T5 span-corruption task, masking 15\% of the input sequence, with an average corrupted span length of 3 tokens.

\noindent \textbf{Unsupervised Contrastive Pre-training:}
In order to align the model with the fine-tuning task, we perform further pre-training following the cropping technique~\citep{DBLP:journals/tmlr/IzacardCHRBJG22}. In this task, given a document, a positive pair ($s$, $s^+$) is sampled by independently cropping two random spans comprising 10 to 50\% of the input. The model is trained to minimize the following contrastive loss:


{\scriptsize
\begin{equation}
\label{eq:contrastive-loss}
\mathcal{L} = - \frac{1}{n} \sum_{i} \log \frac{e^{\mathrm{cos}(f(s_i), f(s_i^+))}}{e^{\mathrm{cos}(f(s_i), f(s_i^+))} + \sum_j e^{\mathrm{cos}(f(s_i), f(s^{-}_{ij}))}} \;,
\end{equation}
}


\noindent where each $s_i$ is associated with one positive example $s_i^+$ as per the sampling technique, and negatives $\{s^{-}_{ij}\}$ are sampled in-batch. We use a batch size of 128 leveraging GradCache~\citep{DBLP:conf/rep4nlp/GaoZHC21}, and cross-device negatives across 4 GPUs. The representations $f(.)$ generated by the model are compared using the cosine similarity function.

\noindent \textbf{Supervised Contrastive Fine-tuning:}
We finally fine-tune the model for retrieval in the MS-MARCO dataset for eight epochs. Both the title and body of the documents are used, as this is the default setting for the document retrieval task. We start with ANCE-MaxP negatives~\citep{DBLP:conf/iclr/XiongXLTLBAO21}, refreshing them every two epochs with the model under training. 
We follow the loss introduced in Equation~\ref{eq:contrastive-loss}, leveraging labeled query-document pairs. We sample 9 negatives per query, using a batch size of 128 and in-batch negatives. Moreover, cross-device negatives are considered across 4 GPUs, which totals 5120 documents for each query in the batch.




\section{Experiments}

This section starts by addressing the overall retrieval performance of the T5-2K model. Then, we show the \textit{dwell in the beginning} behavior that is present in the model, investigating each of the training steps to identify its emergence. 

\subsection{Retrieval Performance}

\begin{table}[t!]
\resizebox{\columnwidth}{!}{\begin{tabular}{@{}lccc@{}}
\toprule
             & Size      & MRR@100 & R@100 \\ \midrule

ANCE-MaxP~\citep{DBLP:conf/iclr/XiongXLTLBAO21}    & 125M  & 0.384   & 0.906 \\ 
ADORE~\citep{DBLP:conf/sigir/ZhanM0G0M21}        & 110M  & 0.405   & 0.919 \\ \midrule
ICT~\citep{DBLP:conf/acl/LeeCT19}          & 110M  & 0.396   & 0.882     \\
SEED~\citep{lu-etal-2021-less}         & 110M  & 0.396   & 0.902    \\ \midrule
RepLLaMA~\citep{DBLP:journals/corr/abs-2310-08319}     & 7B    & 0.456   & -     \\ \midrule
T5-2K (ours) & 220M & 0.414   & 0.915 \\ \bottomrule
\end{tabular}}
\caption{Retrieval results on MS-MARCO documents.}
\label{tab:res}
\end{table}

 
Before moving to the study of the positional biases, we look into the overall performance of our model to assess its soundness, considering the official MS-MARCO evaluation metrics (mean reciprocal rank and recall). For reference, Table~\ref{tab:res} contains retrieval results on the MS-MARCO document dataset (development splits), where our model achieves comparable performance to models trained following similar pipelines. The first group references models that do not leverage pre-training tasks, while the ones in the second group incorporate them. Finally, the third group contains a model that also underwent simple fine-tuning, but has 30 times more parameters. Note that other authors have proposed heavily engineered pre-training tasks that do improve results (e.g., COSTA~\citep{DBLP:conf/sigir/MaGZFC22}, Longtriever~\citep{DBLP:conf/emnlp/YangLLS023}, or LLaRA~\citep{DBLP:journals/corr/abs-2312-15503}), but that is out of scope for this work. Appendix~\ref{sec:appendix} provides additional training details.


\subsection{Impact of Relevant Passage Position}




For a subset of the queries in the MS-MARCO dataset (i.e., 1130 queries), we can cross-reference their relevant documents with the MS-MARCO passage collection to identify the relevant information within the document through exact matching. In a first experiment assessing the impact of the position of the relevant passage, we retrieve from the collection 11 times: First, a default run with the documents unchanged, followed by 10 runs where the documents associated with the queries have the relevant passage moved to different positions. For each document, given its length $l_d$ and the length of the relevant passage $l_p$ (both in tokens), we compute 10 sequential and uniform insertion points ($I_i$) for the passage, according to $I_i = (i-1)\frac{l_d-l_p}{9}, i \in \{1,...,10\}$, moving the passage from its original position to each $I_i$.

\begin{figure}[t!]
    \centering
    

    

    \includegraphics[width=\columnwidth]{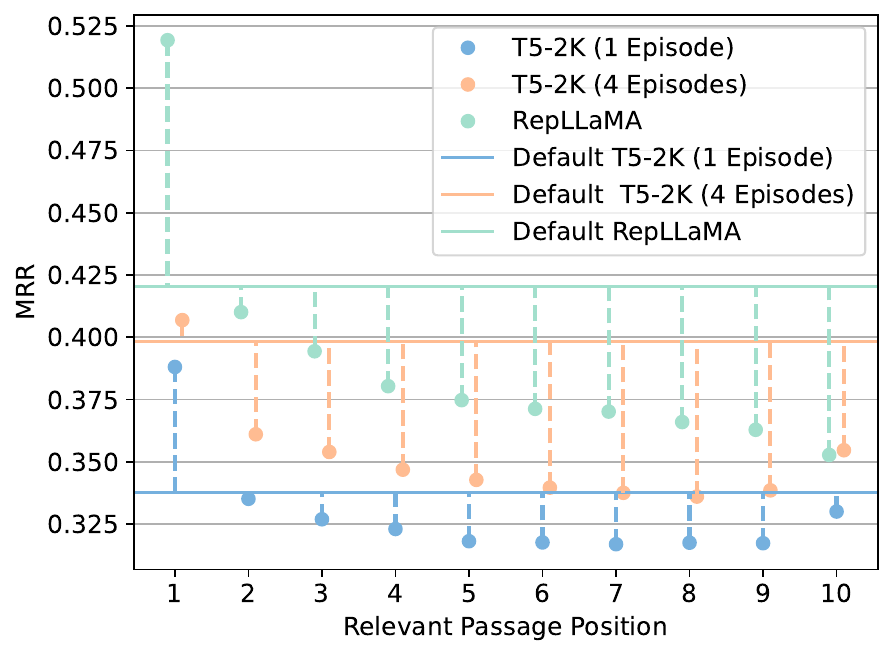}
    \caption{Performance of T5-2K and RepLLaMA. Full lines represent the unchanged version of the documents. Dashed lines represent the variations obtained when the relevant passages are moved to a different position.}
    \label{fig:position}
\end{figure}

The performance of our model after one training episode (i.e., before the first ANCE negative refreshing) is depicted in the blue lines of Figure~\ref{fig:position}. We see that when the relevant passage is moved to the beginning of the document, the performance increases when compared to the default setting (i.e., unchanged documents). Conversely, if the passage is moved anywhere else, the performance drops. The green lines show that the same pattern also holds for RepLLaMA-7B\footnote{\url{https://huggingface.co/castorini/repllama-v1-7b-lora-doc}}~\citep{DBLP:journals/corr/abs-2310-08319}, i.e. a version of LLaMA-2 fine-tuned for dense retrieval on MS-MARCO for one epoch. In other words, a \textit{dwell in the beginning} effect is observed, where the initial positions are heavily preferred to later ones. 

This differs from the \textit{lost in the middle}~\citep{DBLP:journals/corr/abs-2307-03172} phenomena, where performance would drop significantly only in middle sections, rising in the end. We also note that further fine-tuning on MS-MARCO data will aggravate the behavior, as shown by the orange lines in Figure~\ref{fig:position}, given the larger performance mismatch between the default setting and insertion positions other than the first.
\begin{figure}[t!]
    \centering

    \includegraphics[width=0.95\columnwidth, height=3.66cm,]{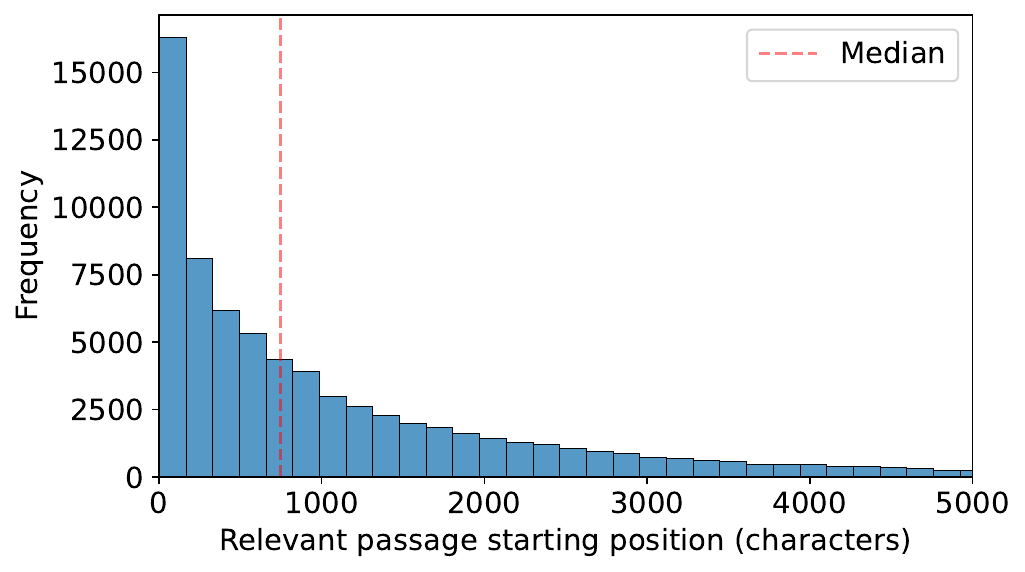}
    
    \caption{Distribution for the starting position (characters) of relevant passages within 75,000 documents from the MS-MARCO training split.}
    \label{fig:distribution-passages}
\end{figure}

To better understand this behavior, we can look at the distribution in Figure~\ref{fig:distribution-passages}, which shows that MS-MARCO documents tend to contain the relevant passage earlier in the document, with the median starting position at 746 characters. This can be impactful for the biases in Figure~\ref{fig:position}, given the lack of examples with relevant information later in the document. To further investigate this phenomenon, the next sub-sections explore the locality of the pre-training tasks to address potential impacts on long-context modeling.


\subsection{Contrastive Pre-training Location Bias}\label{sec:contrastive-bias}

\pgfplotsset{
/pgfplots/custom legend/.style={
legend image code/.code={
\draw [only marks,mark=square*]
plot coordinates { 
(0.3cm,0cm)
};
}, },
}

\begin{figure}[t!]
    \centering
    \includegraphics[width=\columnwidth]{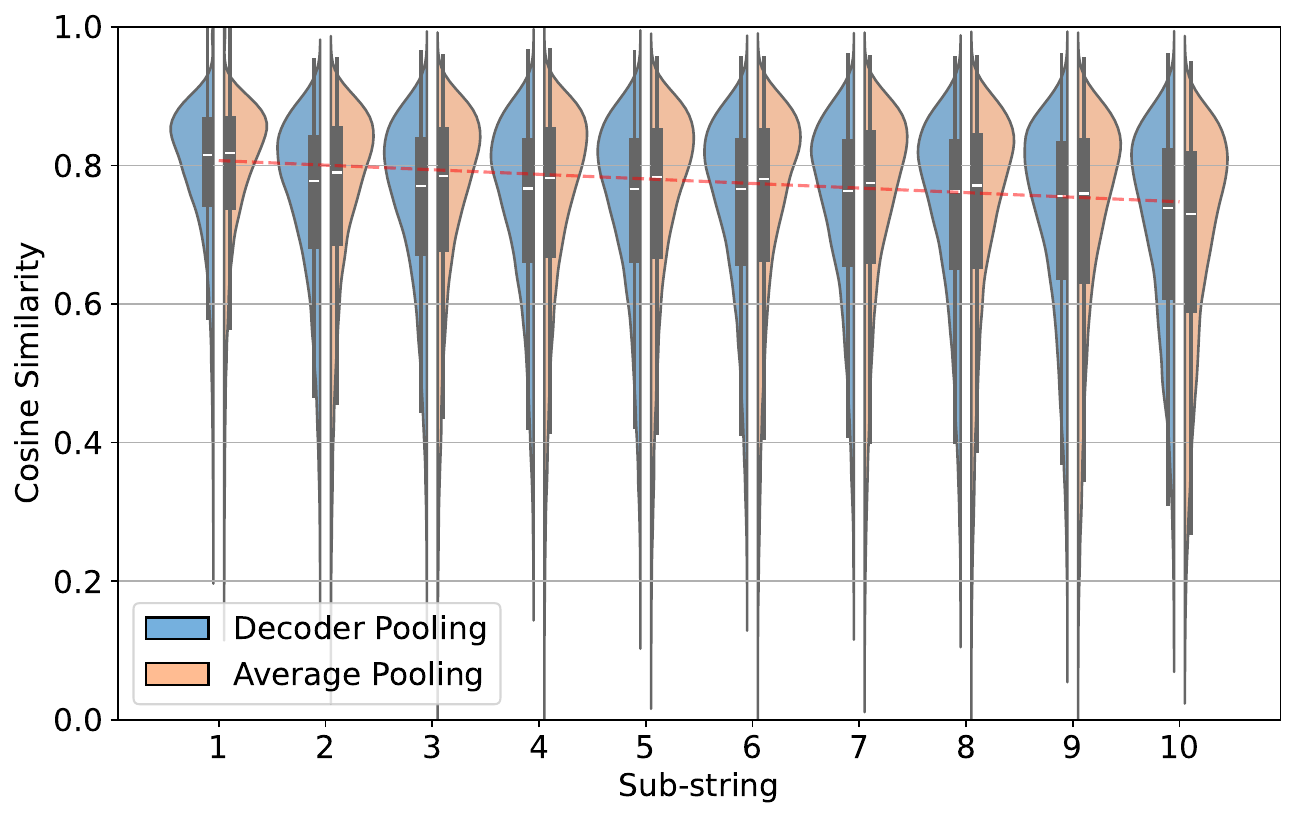}
    \caption{Cosine similarity distribution for exact matching of sub-strings in different locations, using a sample of 24,000 MS-MARCO documents, for the T5-2K  model after contrastive pre-training using both decoder-pooling and average-pooling.}
    \label{fig:distribution-cosine}
\end{figure}

To better estimate positional biases after the contrastive pre-training step, we evaluate the performance of the model on exactly matching sub-strings from different locations. For instance, given a document $d$, 10 sub-strings are sampled by segmenting $d$ in 10 sequential groups with uniform token length. In other words, the first sub-string contains the first 10\% tokens of $d$, while the last sub-string contains the last 10\% tokens. Then, the embedding generated for $d$ is compared with the embedding of each sub-string using the cosine similarity. Figure~\ref{fig:distribution-cosine} shows that the similarity values tend to decrease when the position of the sub-string moves from the beginning, and that this behavior holds for strategies that either use decoder pooling or average pooling of token representations.

This indicates that the representation generated for a document is better at capturing its earlier contents. While in the previous sub-section similar behavior could be justified by the data's underlying distribution, the pseudo-queries and documents for this task were independently sampled from the same uniform distribution over the input. This suggests that the bias is intrinsic to models trained on web documents, e.g. by fitting to information distributions commonly found in real web documents that follow the \textit{inverted pyramid} writing style~\citep{DBLP:journals/corr/abs-1810-09305}, where earlier paragraphs are often more representative. Since web documents are the most common source of contrastive pre-training data~\citep{DBLP:journals/corr/abs-2212-03533, DBLP:journals/tmlr/IzacardCHRBJG22}, this is problematic for tasks where the whole input must be accurately captured, as is for instance the case of retrieval augmented generation~\citep{DBLP:conf/emnlp/ChevalierWAC23, DBLP:journals/corr/abs-2304-08467}.


\subsection{Span Corruption Location Bias}

\begin{figure}[t!]
    \centering
    \begin{tikzpicture}
      \begin{axis}[
        xlabel=Token window,
        ylabel=Accuracy,
        ymin=0.2, ymax=1.0,
        xmin=-1, xmax=10,
        xtick={0.25, 1.25, 2.25, 3.25, 4.25, 5.25, 6.25, 7.25, 8.25, 9.25 },
        xticklabels={1, 2, 3, 4, 5, 6, 7, 8, 9, 10},
        legend pos=south east,
        legend style={font=\tiny},
        width=\columnwidth,
        height=5.7cm,
        tick label style={font=\tiny},
        label style={font=\tiny},
        ymajorgrids
        ]
        
        \addplot+[mark=*,beatpurple, mark options={color=beatpurple}, error bars/.cd,y dir=both,y explicit,] coordinates {
            (0, 0.7482373678025852)   +- (0, 0.23282382374147106)
            (1, 0.7512091394262842)   +- (1, 0.23401479595771826)
            (2, 0.7531336480546964)   +- (2, 0.23345694557486485)
            (3, 0.756491895468078)   +- (3, 0.23331242719868053)
            (4, 0.7536444369973191)   +- (4, 0.2309830034785661)
            (5, 0.7548076923076923)   +- (5, 0.23230999600579766)
            (6, 0.756527523698653)   +- (6, 0.2326591681811125)
            (7, 0.7556302657161373)   +- (7, 0.2354128168538494)
            (8, 0.7509928843289757)   +- (8, 0.2336578733318183)
            (9, 0.7390662964248524)   +- (9, 0.2381768310724322)
        };

        \addplot+[mark=*,SeaGreen, mark options={color=SeaGreen}, error bars/.cd,y dir=both,y explicit
            ] coordinates {
            (0.5, 0.613905064407976)   +- (0, 0.2695659009597244)
            (1.5, 0.629022166607079)   +- (1, 0.2648050506826437)
            (2.5, 0.6354640980735552)   +- (2, 0.2645216718168184)
            (3.5, 0.6392979452054794)   +- (3, 0.2626100831246779)
            (4.5, 0.6459755030621173)   +- (4, 0.262975301502413)
            (5.5, 0.6481707317073171)   +- (5, 0.2609697466985774)
            (6.5, 0.6442506906077348)   +- (6, 0.2586705874314815)
            (7.5, 0.6557939159486549)   +- (7, 0.26125444505926404)
            (8.5, 0.6502476123098692)   +- (8, 0.257168463943617)
            (9.5, 0.645009926006136)   +- (9, 0.26021010826440405)
        };
        
        \legend{T5-rope, T5-base}
        
      \end{axis}
    \end{tikzpicture}

    \caption{Span prediction accuracy on different zones of the input, using 7000 random 3-token spans per window.}
    \label{fig:t5-lm-accuracy}
\end{figure}
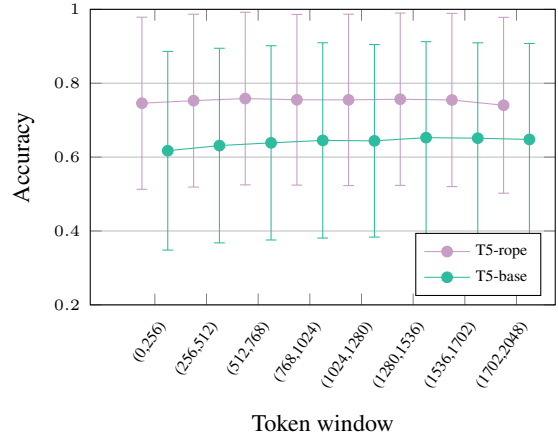

\begin{figure}[t!]
    \centering
    \includegraphics[width=\columnwidth]{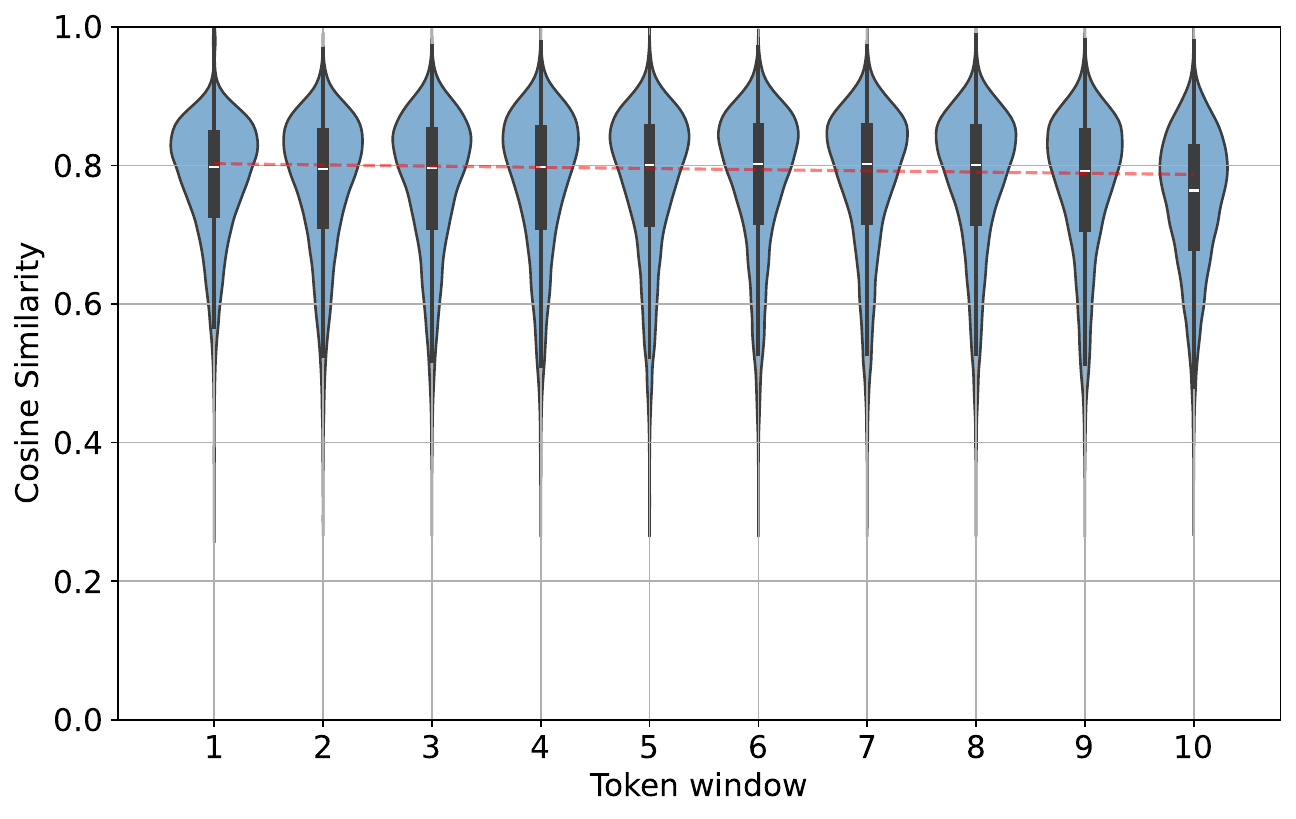}
    \caption{Cosine similarity distribution for exact matching of sub-strings in different locations, using a sample of 24,000 MS-MARCO documents, for the T5-2K model after language model pre-training.}
    \label{fig:distribution-cosine-lm}
\end{figure}

Finally, we look into the language model pre-training task. We evaluate on the original task, by independently corrupting spans of 3 tokens across multiple parts of the input, divided in ten windows as per the previous experiments. Through this, we can see if the accuracy of the model varies when predicting the correct spans across the different parts of the input document.

Figure~\ref{fig:t5-lm-accuracy} shows uniform performance, suggesting no inherent bias in this task using RoPE. We also evaluate the original T5-base, and see that although it shows a slightly higher performance on predicting later positions, it is still rather uniform. As none of the models display the \textit{dwell in the beginning} effect, we conclude that the language modeling pre-training task did not induce any biases, and that this behavior emerged as soon as the embedding task was added to the training pipeline. To further solidify this result, Figure~\ref{fig:distribution-cosine-lm} shows the evaluation of the T5-2K model after language model pre-training (but before embedding-based learning) on the embedding task from Section~\ref{sec:contrastive-bias}, showing a similar pattern to Figure~\ref{fig:t5-lm-accuracy}, without a noticeable \textit{dwell in the beginning} effect.

\section{Conclusions and Future Work}

This study investigated a \textit{dwell in the beginning} effect on Transformer-based models for document retrieval. Through experiments with a T5 model and RepLLaMA, we observed that the embeddings tend to favor information located at the beginning of the input, leading to decreased performance when relevant information is elsewhere in the document. We investigate each step in the training pipeline, namely language model pre-training, contrastive pre-training, and contrastive fine-tuning, showing that biases emerge in the contrastive pre-training step, and that they persist throughout the fine-tuning process. Our findings emphasize the importance of considering the quality of embeddings for long inputs, particularly in contexts where effectively capturing the entire sequence is essential for the downstream task. Moreover, our results can further justify previous research which showed limited gains on long-sequence modeling for MS-MARCO, when compared to aggregation approaches~\citep{DBLP:journals/corr/abs-2207-01262}.

As for future work, we note that while our experiments focused on tied encoders, a similar study can be conducted using untied weights, given the size mismatch between queries and documents. Furthermore, addressing the identified biases may involve devising more robust pre-training tasks, or curating better-distributed datasets, all while considering evaluation on appropriate retrieval benchmarks that require long-context modeling~\citep{DBLP:journals/corr/abs-2305-13915, loco-benchmark}.


\section*{Limitations and Ethical Considerations}

All the datasets and models used in our experiments are publicly available, and we provide the source code that allows for reproduction of the results, as well as model checkpoints.

By using large pre-trained language models, we acknowledge the risks associated with the presence of inherent biases embedded within the models, which may inadvertently perpetuate or amplify societal biases present in the training data.

One limitation in the work reported on this paper relates to the fact that our tests have only used English data. Other languages can expose different phenomena in terms of how document-context is handled, and future work can perhaps consider other datasets such as the one from the NeuCLIR competition~\citep{DBLP:journals/corr/abs-2404-08071}. Doing a similar analysis on other domains besides web documents would also be interesting, and we encourage the research community to further study document-context modeling in connection to different types of information retrieval tasks.

\section*{Acknowledgements}

We thank the anonymous reviewers for their valuable comments and suggestions. This research was supported by the Portuguese Recovery and Resilience Plan through project C645008882-00000055 (i.e., the Center For Responsible AI), and also by the Fundação para a Ciência e Tecnologia (FCT), specifically through the project with reference UIDB/50021/2020 (DOI: 10.54499/UIDB/50021/2020), the project with reference UIDP/04516/2020 (DOI: 10.54499/UIDB/04516/2020), and also through the Ph.D. scholarship with reference PRT/BD/153683/2021 under the Carnegie Mellon Portugal Program. 

\bibliography{acl2023}
\bibliographystyle{acl_natbib}

\appendix

\section{Training Details}
\label{sec:appendix}
This appendix starts by detailing the training setup used in our experiments, and it then presents experimental results that further assess the impact of the different training stages.
\subsection{Hyperparameters}
\label{sec:appendix-hyperparams}

The following subsections detail the hyperparameters used for model training. If a certain element is not stated, the default value from the HuggingFace Trainer API was used. All models were trained in the same computational infrastructure with 4 NVIDIA A100 40GB GPUs.

\subsubsection{Span Corruption Pre-training}

\begin{table}[h!]
\centering
\begin{tabular}{@{}l@{\hspace{2.4cm}}l@{}}
\toprule
Optimizer             & AdamW            \\
Initial learning rate & 1e-5             \\
Scheduler             & Cosine           \\
Batch size            & 80               \\
Gradient accumulation & 16               \\
Gradient clipping     & 1                \\
Weight decay          & 0                \\
Total steps           & 49152            \\
Warm-up steps         & 10\%             \\ \bottomrule
\end{tabular}
\caption{Set of hyperparameters considered for span-corruption pre-training.}
\end{table}

\newpage
\subsubsection{Contrastive Pre-training}

\begin{table}[h!]
\centering
\begin{tabular}{@{}l@{\hspace{2cm}}l@{}}
\toprule
Optimizer             & AdamW            \\
Initial learning rate & 5e-6             \\
Scheduler             & Linear \\
Batch size            & 128               \\
Gradient accumulation & 1               \\
Gradient cache chunk size & 24           \\
Hard negatives per query & 0           \\
Epochs & 1           \\
\bottomrule
\end{tabular}
\caption{Set of hyperparameters considered for contrastive pre-training.}
\end{table}

\subsubsection{Fine-tuning}

\begin{table}[h!]
\centering
\begin{tabular}{@{}l@{\hspace{2cm}}l@{}}
\toprule
Optimizer             & AdamW            \\
Initial learning rate & 5e-6             \\
Scheduler             & Linear \\
Batch size            & 128               \\
Gradient accumulation & 1               \\
Gradient cache chunk size & 24           \\
Hard negatives per query & 9           \\
Epochs & 8           \\
\bottomrule
\end{tabular}
\caption{Set of hyperparameters considered for final model fine-tuning.}
\end{table}

\subsection{Impact of Each Training Step}
\label{sec:appendix-pipeline}

Table~\ref{tab:abl} aligns our training pipeline with previous work, showing the importance of the pre-training tasks, and the benefits of multiple fine-tuning steps with negative refreshing. Note that the performance without any pre-training is particularly low since the model had no previous exposure to the new rotary embeddings.

\begin{table}[ht!]

\resizebox{\columnwidth}{!}{\begin{tabular}{@{}cccll@{}}
\toprule
\multicolumn{1}{l}{\begin{tabular}[c]{@{}l@{}}LM \\ Pre-training\end{tabular}} & \multicolumn{1}{l}{\begin{tabular}[c]{@{}l@{}}Contrastive \\ Pre-training\end{tabular}} & \multicolumn{1}{l}{Fine-tuning} & MRR     & R@100   \\ \midrule
\xmark                                                         & \xmark                                                                   & 1 episode                          & 0.177 & 0.632 \\
\cmark                                                     & \xmark                                                                   & 1 episode                          & 0.350   & 0.872   \\
\cmark                                                     & \cmark                                                              & 1 episode                          & 0.372   & 0.889   \\ \midrule
\cmark                                                     & \cmark                                                              & 4 episodes                          & 0.414   & 0.915   \\ \bottomrule
\end{tabular}}
\caption{Performance on MS-MARCO for different combinations of pre-training tasks, and after fine-tuning.}
\label{tab:abl}
\end{table}

\end{document}